\documentclass[aps,pre,preprint]{revtex4-1}

\usepackage{graphicx}% Include figure files
\usepackage{subfig}
\usepackage{subfloat}
\usepackage{dcolumn}% Align table columns on decimal point
\usepackage{bm}% bold math
\usepackage[utf8]{inputenc}
\usepackage[T1]{fontenc}
\usepackage{mathptmx}
\usepackage{amsmath}
\usepackage{etoolbox}
\usepackage{hyperref}
\usepackage{xcolor}

\draft % marks overfull lines with a black rule on the right

\begin{document}

%\preprint{}

\title{Classical density functional theory for nanoparticle-laden droplets} 
%\title{Nanoparticle-laden droplets within cDFT} 


\author{Melih G\"ul}
\email[]{melih.guel@uni-tuebingen.de}
\affiliation{ Institute for Theoretical Physics, University of Tübingen, Auf der Morgenstelle 14, 72076 Tübingen, Germany}

\author{A.~J.~Archer}
\email{a.j.archer@lboro.ac.uk}
 \affiliation{Department of Mathematical Sciences and Interdisciplinary Centre for Mathematical Modelling, Loughborough University, Loughborough LE11 3TU, UK}%Lines break automatically or can be forced with \\
 
\author{B.~D.~Goddard}%
 \email{b.goddard@ed.ac.uk}
\affiliation{School of Mathematics and the Maxwell Institute for Mathematical Sciences, University of
Edinburgh, Edinburgh EH9 3FD, UK}%

\author{Roland Roth}
\email[]{roland.roth@uni-tuebingen.de}
\affiliation{Institute for Theoretical Physics, University of Tübingen, Auf der Morgenstelle 14, 72076 Tübingen, Germany}
%\thanks{}
%\altaffiliation{Institut f\"ur Theoretische Physik T\"ubingen}
%\affiliation{Eberhard-Karls-Universitaet Tuebingen}


\date{\today}

\begin{abstract}
Droplets of a pure fluid, such as water, in an open container surrounded by gas, are thermodynamically unstable and evaporate quickly. In a recent paper [Archer et al.\ J.\ Chem.\ Phys.\ {\bf 159}, 194403 (2023)] we employed lattice density functional theory (DFT) to demonstrate that nanoparticles or solutes dissolved in a liquid droplet can make it thermodynamically stable against evaporation. In this study, we extend our model by using continuum DFT, which allows for a more accurate description of the fluid and nanoparticle density distributions within the droplet and enables us to consider size ratios between nanoparticles and solvent particles up to 10:1. While the results of the continuum DFT agrees well with those of our earlier lattice DFT findings, our approach here allows us to refine our understanding of the stability and structure of nanoparticle laden droplets. This is particularly relevant in light of the recent global COVID-19 pandemic, which has underscored the critical role of aerosol particles in virus transmission. Understanding the stability and lifetime of these viron-laden aerosols is crucial for assessing their impact on airborne disease spread.
%, which is relevant in the context of viron laden aerosol particles that can %stay airborne for a long time.
\end{abstract}


\maketitle %\maketitle must follow title, authors, abstract and \pacs
\newpage
\section{Introduction}

Airborne droplets produced, for example, through people coughing can last for a surprisingly long time in the air depending on their size, humidity of the air, and if non-volatile particles are in the droplet \cite{wells1934,netz2020mechanisms,Archer2023}. The lifetime of a droplet is prescribed by the competition between gravity and evaporation, hence large droplets fall down to the ground within a few seconds whereas smaller droplets evaporate rapidly, especially in environments with low humidity. As discussed in Refs.~\cite{wells1934,netz2020mechanisms}, for a large droplet the sedimentation time, which scales as $1/R^2$, where $R$ is the radius of the droplet, is much smaller than the evaporation time, while the situation is reversed in the case of small droplets.

These aerosol droplets play a crucial role in spreading suspended viruses or other materials over a long distance, if they remain airborne for a long time. This is also relevant to the dispersal into the environment and onto livestock of the various chemicals used in crop sprays \cite{felsot2010agrochemical}. The suspended particles can stabilize an otherwise evaporating droplet and hence prolong the droplet's lifetime in air. This effect has far reaching repercussions on the spread of infections, where viruses within water droplets close to the surface enhance the contagion for a long period of time.
The reason why a pure droplet evaporates lies in the fact that for a convexly shaped body such as a droplet, the pressure inside and outside the droplet differ by the Laplace pressure, which is given by twice the surface tension divided by its radius. Hence, due to this imbalance, the droplet eventually evaporates. Adding non-volatile particles, such as nanoparticles, into the droplet can prevent it from reaching complete evaporation, since attractive forces between these nanoparticles and the liquid reduce the pressure inside the droplet, so that mechanical equilibrium can be reached \cite{wells1934,netz2020mechanisms,Archer2023}.

We investigate such nanoparticle laden droplets in the framework of classical density functional theory (DFT) \cite{evans1979nature, hansen2013theory} \textcolor{black}{and demonstrate that these nanoparticles are able to stabilize the droplet from complete evaporation}. We treat the laden droplet as a binary mixture of hard-sphere particles with square-well attractions between them, also allowing for different diameters of the components. The hard-sphere part of the associated Helmholtz free energy functional is well described by fundamental measure theory (FMT) \cite{rosenfeld1989,roth2010fundamental}, however with deficiencies when it comes to highly asymmetric mixtures of hard spheres \cite{roth2010fundamental} where the free energy cannot be fully captured by FMT. Our treatment of the attractive square-well uses the somewhat crude random phase approximation (RPA) or mean-field approximation \cite{hansen2013theory,archer2017,Melih2025effective}.

The lattice-DFT calculations of Ref.~\cite{Archer2023} give a limited description of the density distributions and thus the structure of the particles within the equilibrium droplets, due to the coarse-graining approximation of treating the system as discretized onto a lattice. In contrast, the continuum DFT calculations presented in this work give much more insight regarding how the particles are ordered within the interior of the droplet and at the interface, describing well, e.g., the adsorption of solute particles close to the surface. We find that both the solvent and solute particles tend to coordinate themselves in concentric shells, leading to oscillatory density distributions within the droplets. This is perhaps not too surprising, given the strong spherically symmetric confinement within the droplet. As well as the accumulation of liquid and nanoparticles that can be captured by this approach, it can also include the influence of size disparity between the components \cite{fortini2016dynamic,he2021dynamical,kundu2022dynamic,Melih2025effective,lee1985,abe2014}. Indeed, we allow for a size ratio between the liquid and nanoparticles, treating the latter as being between two to ten times the size of the former. While non-volatile particles suspended in water droplets can have a much larger size ratios up to roughly one thousand, our choice is limited by numerical feasibility and accuracy of FMT. \textcolor{black}{Furthermore, we are not including any associative interactions which otherwise would be necessary to try to capture the properties of a water droplet. Nonetheless, we think that stabilization of droplets through the presence nanoparticles is not restricted to specific properties of a water droplet, but is an equilibrium mechanism applicable to any fluid.}
DFT has been used previously to study droplets in one-component systems, for example, to estimate nucleation rates \cite{oxtoby1988nonclassical, archer2011nucleation} and interfacial properties \cite{malijevsky2012perspective}.
The work of \cite{liwu2008} investigates the properties of very small droplets of a Lennard-Jones fluid in a super saturated environment leading to vapor-liquid nucleation. They found that DFT accurately predicted the microscopic structure, size and free-energy barrier of critical nuclei. \textcolor{black}{There are Molecular Dynamics (MD) simulations on the time evolution of nanoparticle laden droplets with similar setups \cite{jin2023morphology,zhan2022molecular,li2013cloud,chen2013molecular} providing information about the morphological structure of droplets and the decrease of their diameters during evaporation. The MD calculations enable to investigate, in particular, the structure of the droplets and other quantities describing their evaporation.}

Alongside DFT, we employ a simple model for stable droplets containing solute particles, which we also refer to as nanoparticles, that can predict the size of a droplet for given number of solutes and humidity of the surrounding air. The underlying ideas rely on the conditions of mechanical and chemical equilibrium, the former taking into account the surface tension of the droplet and the latter incorporating humidity of the ambient air where additionally the solvent liquid particles are exchanged between the droplet and its environment.
The corresponding results of the thermodynamic model and the results stemming from our DFT calculations can be compared in order to solidify our understanding of the nanoparticle laden droplet as a binary mixture.
The study in Ref.~\cite{Archer2023} compared the lattice-DFT results with those derived from a capillarity-approximation based thermodynamic model. It also used dynamical density functional theory (DDFT) to further study the evaporation and condensation dynamics of the liquid, showing stability and formation dynamics of droplets \cite{hansen2013theory,marconi1999dynamic,archer2006dynamical,te2020classical,kierlik2001capillary}. These lattice-type DFT calculations in different variants were also employed in studies of liquid droplets on surfaces \cite{hughes2014introduction,hughes2015liquid,chalmers2017dynamical,areshi2019kinetic} and liquids adsorbed in pores and porous media \cite{kierlik2001capillary,woo2001mean,schneider2014filling}. The improved approach for nearest-neighbor interactions of \cite{maeritz2021density,maeritz2021droplet} could give an even more precise description of density distribution of liquid and nanoparticles.

This paper is organized as follows: Firstly, we begin by introducing essential parts of the theory in Sec.~\ref{sc::theory}. The square-well potential used to model the attractive forces between solvent and solute particles is outlined in Sec.~\ref{ssc::sw-fluid} and treated within the framework of classical DFT, with emphasis on having bulk phase coexistence between the pure liquid and vapor phases. In Sec.~\ref{ssc::laden-droplet} we generalize to square-well binary mixtures, where we make use of the powerful framework of FMT for hard-sphere mixtures, together with the standard mean-field approximation for the contribution to the Helmholtz free energy stemming from the square-well attractive interactions. We then present the procedure for minimization of the free energy for the binary mixture. This necessitates several steps of pre-minimization, due to numerical challenges that arise for asymmetric mixtures. A simple thermodynamic model is introduced in Sec.~\ref{ssc::Model-Equ}, for a system consisting of a droplet of liquid laden with nanoparticles or solutes surrounded by vapor and, depending on the parameters at hand, partly by solutes. We furthermore relate our theoretical description to the capillarity model discussed in \cite{Archer2023}. Thereafter, we present in Sec.~\ref{sc::results} droplet density profiles at equilibrium for binary mixtures of 2:1 and 10:1 solute to solvent size ratios; the latter ratio is closer to the case of viruses suspended in water droplets. We consider several values of the humidity in order to analyze its influence on the size and density distribution of the droplet. In addition, we compare our model results of Sec.~\ref{ssc::Model-Equ} with the results of our DFT calculations, e.g.\ by calculating the total amount of liquid particles in the droplet. Lastly, we make some concluding remarks in Sec.~\ref{sc::conclusion} on our findings and analysis of particle-laden droplets and possible further work on this.    

\section{Theory}\label{sc::theory}
\subsection{Square-Well Fluid}\label{ssc::sw-fluid}
The one-component square-well (SW) fluid is a fairly simple model for systems of interacting particles with isotropic attractive forces, i.e.\ those with little to no angular dependence. The pair-interaction potential consists of a hard-core repulsion of diameter $\sigma$ and a short-ranged attraction 
\begin{equation}\label{eq::square-well-potential}
	\phi(r)=\begin{cases}
		\infty,\quad r<\sigma,\\
		-\epsilon,\quad \sigma\leq r\leq\lambda\sigma,\\
		0,\quad \text{otherwise},
	\end{cases}
\end{equation}
where $\epsilon$ defines the attraction strength and $\lambda$ the range of the interaction. Here, we employ classical DFT which is a powerful framework to access structure and thermodynamics of a fluid \cite{evans1979nature, hansen2013theory}. By minimizing the one-component grand potential functional
\begin{equation}
	\Omega[\rho]={\cal F}[\rho] + \int \rho(\mathbf{r})\left(V_\text{ext}(\mathbf{r})-\mu\right)d\mathbf{r}
\end{equation}
with respect to the one-body density $\rho(\mathbf{r})$
\begin{equation}\label{eq::DFT-minimization}
	\frac{\delta\Omega[\rho]}{\delta\rho(\mathbf{r})}\bigg|_{\rho(\mathbf{r})=\rho_\text{eq}(\mathbf{r})}=0,
\end{equation}
we obtain the equilibrium density distribution $\rho_\text{eq}(\mathbf{r})$ for the pure solvent. $V_\text{ext}(\mathbf{r})$ is the external potential acting on the fluid, $\mu$ is the chemical potential and ${\cal F}[\rho]={\cal F}_\text{id}[\rho]+{\cal F}_\text{ex}[\rho]$ the intrinsic Helmholtz free-energy functional that can be split into an exactly known ideal-gas part
\begin{equation}\label{eq::ideal-functional}
	{\cal F}_\text{id}[\rho]=\int \rho(\mathbf{r})\left(\log\left(\Lambda^3\rho(\mathbf{r})\right)-1\right)d\mathbf{r},
\end{equation}
where $\Lambda$ is the thermal de Broglie wavelength of the particles, and an excess part containing all the information of the particle interactions. We consider the mean-field functional to describe the SW fluid \cite{hansen2013theory, archer2017}
\begin{equation}\label{eq::excess-functional}
	{\cal F}_\text{ex}[\rho]={\cal F}_\text{hs}[\rho]+\frac{1}{2}\int\int \rho(\mathbf{r})\rho(\mathbf{r}')\phi_\text{sw}(|\mathbf{r}-\mathbf{r}'|)d\mathbf{r}\, d\mathbf{r}',
\end{equation}
where ${\cal F}_\text{hs}[\rho]$ accounts for the hard-sphere repulsion well described by the White-Bear (WB) functional \cite{Roth02,yuwu2002} of FMT \cite{rosenfeld1989, roth2010fundamental}. By extending the attractive part to the inside of the core, i.e. $\phi_\text{sw}(r)=-\epsilon\Theta(\lambda\sigma-r)$, we can compensate the underestimation of correlations \cite{hansen2013theory,archer2017}.

For a (uniform) bulk fluid with bulk density $\rho_b$ and corresponding packing fraction $\eta=\frac{\pi}{6}\rho_b\sigma^3$, Eq.~\eqref{eq::excess-functional} gives the excess free-energy density
\begin{equation}\label{eq::free-energy-density}
	\beta f_\text{ex}=\frac{\beta{\cal F}_\text{ex}[\rho_b]}{V}=\rho_b\frac{4\eta-3\eta^2}{(1-\eta)^2}-4\beta\epsilon\rho_b\eta\lambda^3,
\end{equation}
where $\beta=1/(k_BT)$ and where $k_B$ is Boltzmann's constant and $T$ is the temperature. The corresponding equation of state is
\begin{equation}\label{eq::SW-EoS}
	\beta P = \rho_b\frac{1+\eta+\eta^2-\eta^3}{(1-\eta)^3}-4\beta\epsilon\rho_b\eta\lambda^3,
\end{equation}
where $P$ is the pressure and chemical potential
\begin{equation}\label{eq::SW-mu}
	\beta\mu = \text{const.}+\log(\eta)+\frac{\eta(8-9\eta+3\eta^2)}{(1-\eta)^3}-8\beta\epsilon\lambda^3\eta.
\end{equation}
The SW fluid gives rise to gas-liquid phase separation, i.e.\ for given vapor and liquid densities, $\rho_v$ and $\rho_l$ respectively, we have phase coexistence when
\begin{equation}\label{eq::phase-coexistence}
	P(\rho_v) = P(\rho_l),\quad \mu(\rho_v) = \mu(\rho_l),
\end{equation}
as long as we are below the critical temperature $T_c$. In this paper, we employ values for the one-component SW fluid $\beta\epsilon=1.2$ and $\lambda=1.5$ with the following corresponding liquid and vapor packing fractions at bulk phase coexistence
\begin{equation}\label{eq::coex-etas}
	\eta_l = 0.380225,\quad \eta_v=0.005861
\end{equation}
and a (reduced) surface tension $\tilde{\gamma}=\beta\gamma\sigma^2=0.25$. These are considered here as reference values of the ``liquid'' component which is regarded as the solvent of the nanoparticles in the upcoming discussion.

\subsection{Nanoparticle-Laden Droplets as Binary SW Mixtures}\label{ssc::laden-droplet}

We model a nanoparticle laden droplet as a SW binary mixture within DFT by employing spherical symmetry. Furthermore, we refer to the first component as the ``liquid'' or ``solvent'' with density $\rho_l(\mathbf{r})$ and the second component as the ``nanoparticles'' with density $\rho_n(\mathbf{r})$.

Then, it is straightforward to extend the excess functional in Eq.~\eqref{eq::excess-functional} to a binary mixture as follows,
\begin{gather}\label{eq::excess-functional-binary}
	{\cal F}_\text{ex}[\rho_l,\rho_n] = {\cal F}_\text{hs}[\rho_l,\rho_n]+\frac{1}{2}\int\int \rho_l(\mathbf{r})\rho_l(\mathbf{r}')\phi^{(ll)}_\text{sw}(|\mathbf{r}-\mathbf{r}'|)d\mathbf{r}\, d\mathbf{r}'+\\\nonumber
	+\frac{1}{2}\int\int \rho_n(\mathbf{r})\rho_n(\mathbf{r}')\phi^{(nn)}_\text{sw}(|\mathbf{r}-\mathbf{r}'|)d\mathbf{r}\, d\mathbf{r}'
    +\int\int \rho_l(\mathbf{r})\rho_n(\mathbf{r}')\phi^{(ln)}_\text{sw}(|\mathbf{r}-\mathbf{r}'|)d\mathbf{r}\, d\mathbf{r}',
\end{gather}
where we now distinguish between the attractive SW interactions between the two species as liquid-liquid $\phi^{(ll)}_\text{sw}$, nano-nano $\phi^{(nn)}_\text{sw}$ and inter-component liquid-nano $\phi^{(ln)}_\text{sw}$ with energies $\epsilon_{ll}$, $\epsilon_{nn}$, $\epsilon_{ln}$ and ranges $\lambda_{ll}$, $\lambda_{nn}$, $\lambda_{ln}$, respectively. These SW parameters can in general be chosen independently. However, certain mixing rules such as Lorentz-Berthelot prescribe the inter-component interactions of a SW mixture, see \cite{Martinez2016, Heyes1992, Melih2025effective}.

We minimize the grand functional $\Omega[\rho_l,\rho_n]$ with respect to $\rho_l$ and $\rho_n$
\begin{align}\label{eq::ELGs-binary}
	\frac{\delta\Omega[\rho_l,\rho_n]}{\delta\rho_l(\mathbf{r})}&=0\\\nonumber
	\frac{\delta\Omega[\rho_l,\rho_n]}{\delta\rho_n(\mathbf{r})}&=0,\quad N_n=\text{const.},
\end{align}
where we use the restriction on $\rho_n$ that the number of nanoparticles $N_n=\int d\mathbf{r}\,\rho_n(\mathbf{r})$ is constant. This means that we are describing the binary mixture in the semi-grand canonical ensemble, i.e.\ the liquid is treated grand canonically and the non-volatile nanoparticles are treated canonically.

From Eq.~\eqref{eq::ELGs-binary} we obtain in the absence of any external potentials ($V_{ext}(\mathbf{r})=0$) the implicit equations for $\rho_l$ and $\rho_n$
\begin{align}\label{eq::ELGs-binary-2}
	\rho_l(\mathbf{r}) &= \rho_l^{(0)}\exp\left(c_l^{(1)}(\mathbf{r})+\beta\mu_l\right)\\\nonumber
	\rho_n(\mathbf{r})&=\rho_n^{(0)}\exp\left(c_n^{(1)}(\mathbf{r})\right)
\end{align}
with the one-body correlation functions $c_i^{(1)}(\mathbf{r})\equiv-\delta\beta{\cal F}_{ex}[\rho_l,\rho_n]/\delta\rho_i(\mathbf{r}),\,i = l,n$ and the excess chemical potential of the liquid $\mu_l$. We want to emphasize that since the nanoparticles are treated canonically there is no chemical potential in the exponential of Eq.~\eqref{eq::ELGs-binary-2} for $\rho_n$.

The reason for treating the nanoparticles in the canonical ensemble is that these are not at all volatile. Moreover, they are fairly strongly bound to the water within the droplets and so when sampling typical configurations of the system, these should not be placed far outside of the droplet, as would be the case if we treated the nanoparticles in the grand canonical ensemble. In contrast, by treating the liquid in the grand canonical ensemble, we can easily and naturally specify the humidity conditions of the surrounding vapor. Thus, we can investigate nanoparticle-laden droplets in different environments; the relative humidity will impact on the equilibrium size and density of such droplets.

We solve Eq.~\eqref{eq::ELGs-binary-2} iteratively using the Picard iteration scheme, where we mix the solution of the $k$-th step $\rho_i^{(k)}$, where $i=l,n$, with the right-hand side (rhs) of Eq.~\eqref{eq::ELGs-binary-2}, $\rho_i^{(rhs,k)}$
\begin{equation}\label{eq::Picard-iteration}
	\rho_{i}^{(k+1)}(\mathbf{r})=(1-\alpha_i)\rho_{i}^{(k)}(\mathbf{r})+\alpha_i\rho_{i}^{(\text{rhs},k)}(\mathbf{r}).
\end{equation}
Here, $\alpha_i$ are mixing parameters that can be chosen independently from each other. Indeed, it is typical for $\alpha_n$ to be much smaller than $\alpha_l$. Eq.~\eqref{eq::Picard-iteration} is repeated until the densities $\rho_i$ have converged to within a given tolerance.

As it turns out, the Picard scheme given in Eq.~\eqref{eq::Picard-iteration} reaches its limit of applicability for a binary mixture of high size ratio, i.e.\ it struggles when one component is much bigger than the other. The reason for this lies in the fact that the exponential in Eq.~\eqref{eq::ELGs-binary-2}, $\exp\left(c^{(1)}_n(\mathbf{r})\right)$, becomes very large in this scenario of a highly asymmetric binary mixture, leading to numerical instabilities. Therefore, it is necessary to employ a different kind of minimization scheme where we also make use of several stages of pre-minimization. This allows us to come closer to the real solutions $\rho_l(\mathbf{r})$ and $\rho_n(\mathbf{r})$ thus improving upon numerical stability. The following four stages are successively performed:

\begin{enumerate}
\item We first start considering the second component of nanoparticles with inital density $\rho_n(r)=\rho^{(0)}_n\Theta(R-r)$, where $R$ is the droplet radius in equilibrium and $\rho^{(0)}_n$ the uniform nanoparticle density within the droplet. Then, in order to keep these nanoparticles within the droplet we impose an effective external potential $V_{\text{ext}}(\mathbf{r})$ of the following form
\begin{equation}\label{eq::eff-ext-potential}
    V_{\text{ext}}(r) = V_{\text{in}} + \frac{V_{\text{out}}-V_{\text{in}}}{1+\exp\left(-a(r-R)\right)},\quad V_{\text{in}}=-\epsilon_{ln}\left(\lambda_{ll}q+\lambda_{nn}\right)^3\eta_l^{(0)}\eta_n^{(0)}
\end{equation}
with $a$ and $V_{\text{out}}$ being positive constants that we choose appropriately. $V_\text{in}$ ensures that nanoparticles prefer to stay inside the droplet. Therefore, we parametrize the effective external potential, Eq.~\eqref{eq::eff-ext-potential}, with a sigmoidal function such that the interior of the droplet resembles an attractive region at the exterior a repulsive region for nanoparticles, see Fig.~\ref{effective_potential}.
\begin{figure}
\includegraphics[scale=0.3]{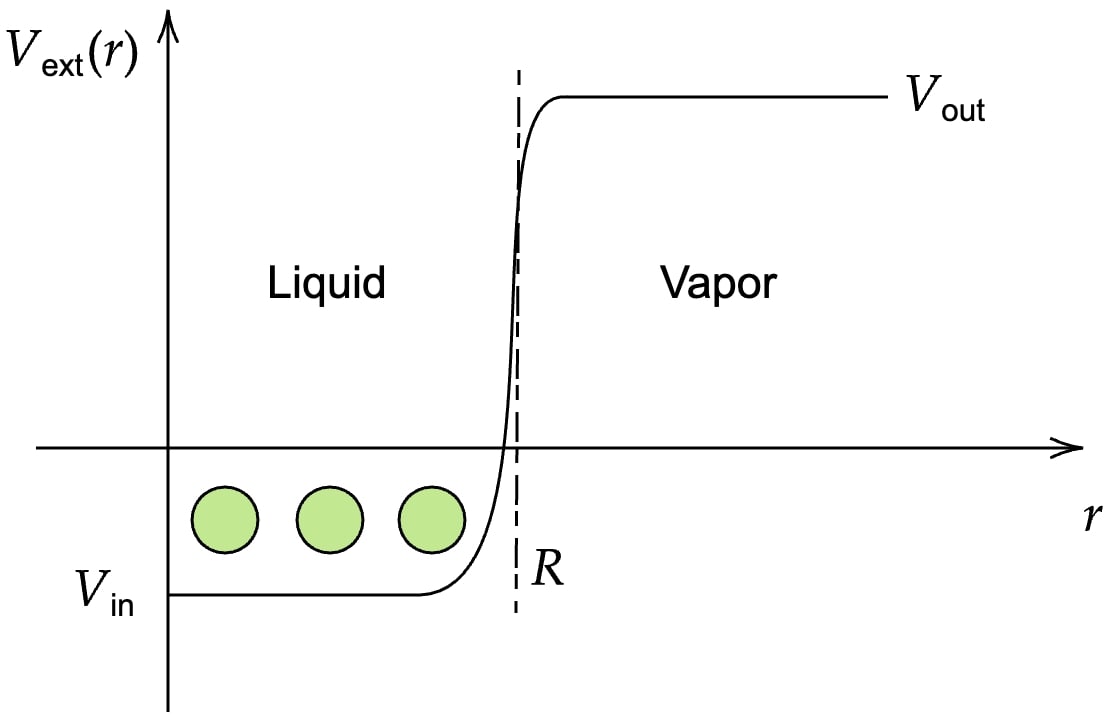}
\caption{The effective external potential $V_\text{ext}(\mathbf{r})$ confines the nanoparticles (green) in a spherical region of radius $R$ resembling the droplet.\label{effective_potential}}
\end{figure}
Here, $\eta_l^{(0)}$ and $\eta_n^{(0)}$ are the uniform packing fractions of the liquid and nanoparticles in equilibrium, respectively and $q=\sigma_l/\sigma_n$ is the size ratio. With this external potential we perform the Picard-scheme, Eq.~\eqref{eq::Picard-iteration}, for the nanoparticle density $\rho_n(\mathbf{r})$
\begin{equation}\label{2-comp-vext}
    \rho_n^{(k)}(r) = (1-\alpha_n)\rho_n^{(k-1)}(r)+\alpha_n\rho_n^{(0)} \exp\left(c^{(1)}_n(r)-\beta V_{\text{ext}}(r)\right),\quad N_n=\text{const.},
\end{equation}
where the mixing parameter $\alpha_n$ is often taken to be 0.1.
It should be emphasized that at this first stage we have not included the liquid density $\rho_l(r)$ and therefore at this stage is not included in $c^{(1)}_n(r)$.

\item Next, we Picard iterate the liquid density profile $\rho_l(r)$ initialized as $\rho_l(r)=\rho_l\Theta(R-r)+\rho_v\Theta(r-R)$ while keeping the nanoparticle density profile $\rho_n(r)$ fixed as a background
\begin{equation}\label{1-comp-iterate}
    \rho_l^{(k)}(r) = (1-\alpha_l)\rho_l^{(k-1)}(r)+\alpha_l\rho_l^{(0)}\exp\left(c^{(1)}_l(r)+\beta\mu_l\right),\quad \rho_n(r)=\text{const,}
\end{equation}
where similarly $\alpha_l=0.1$. The background profile $\rho_n(r)$ from the previous stage enters through the one-body correlation $c^{(1)}_l(r)$.

\item Having obtained the density profiles from the previous two steps, several thousand Picard iterations are performed, as prescribed in Eq.~\eqref{eq::Picard-iteration}. At this point, it is crucial to choose the mixing parameter, $\alpha_n$, as $\alpha_n=\tilde{\alpha}_n\exp\left(-||c^{(1)}_n||_{\text{max}}\right)$ which prevents the exponential in Eq.~\eqref{eq::ELGs-binary-2} from becoming too large. The new mixing parameter $\tilde{\alpha}_n$ is often used in the range $[0.01, 0.1]$.

\item Finally, we make use of the Ng-algorithm, described in the appendix of \cite{Ng1974}, which significantly improves the accuracy of our minimization. For that, several intermediate steps are performed from which mixing parameters are obtained in order to yield the new solution in the iteration.
\end{enumerate}

As we have seen in Sec.~\ref{ssc::sw-fluid}, a SW fluid can coexist in its liquid and vapor phases. Inserting nanoparticles into the SW fluid changes the coexistence densities and chemical potentials without overall altering its qualitative phase behavior, at least for the parameter values we consider in this paper.

The excess bulk free-energy density of the binary mixture can be obtained from Eq.~\eqref{eq::excess-functional-binary} for constant densities $\rho_i$
\begin{equation}\label{eq::free-energy-density-binary}
	\beta f_\text{ex}\left(\rho_l,\rho_n\right)=\beta f^\text{hs}_\text{ex}(\rho_l,\rho_n)+\frac{1}{2}\rho_l^2w_{ll}+\frac{1}{2}\rho_n^2w_{nn}+\rho_l\rho_nw_{ln},
\end{equation}
where we have set $w_{ij}=\int d\mathbf{r}\,\beta\phi^{(ij)}_{sw}(\mathbf{r})$. The free-energy density $f^{hs}_{ex}$ is given by the WB functional \cite{Roth02,yuwu2002} for a binary mixture in the bulk only depending on the densities $\rho_i$. With Eq.~\eqref{eq::free-energy-density-binary} we obtain the pressure $P$ and the chemical potentials $\mu_l$ and $\mu_n$ of the binary mixture
\begin{align}\label{eq::pressure-chemPot-binary}
	\mu_l &=\frac{\partial f_{ex}}{\partial\rho_l},\quad\mu_n =\frac{\partial f_{ex}}{\partial\rho_n}\\\nonumber
	P &= -f_{ex} + \rho_l\mu_l+\rho_n\mu_n.
\end{align}
More specifically, the total pressure $P$ is decomposed into two parts
\begin{equation}\label{eq::total-pressure}
    P = P_\text{BMCSL}(\rho_l, \rho_n)+P_\text{sw}(\rho_l,\rho_n),
\end{equation}
where $P_\text{BMCSL}$ is the Boublik-Mansoori-Carnahan-Starling-Leland (BM-CSL) pressure for a mixture of hard spheres \cite{mansoori1971equilibrium, carnahan1969} and $P_\text{sw}$ is the contribution stemming from the SW interactions and which comes from the SW free-energy density in Eq.~\eqref{eq::free-energy-density-binary}. The chemical potentials in Eq.~\eqref{eq::pressure-chemPot-binary} are split into hard-sphere and SW parts in the same way. For instance, the SW part of the chemical potential of liquid $\mu_l$ reads
\begin{equation}\label{eq::chem-pot-liquid}
    \beta\mu_{l,\text{sw}} = -8\beta\epsilon_{ll}\lambda_{ll}^3\eta_l-8\beta\epsilon_{ln}\lambda_{ln}^3\eta_n,\quad \eta_i=\frac{\pi}{6}\rho_i\sigma_i^3,\quad i = l,n,
\end{equation}
where we have evaluated the $w_{ij}$ explicitly. The chemical potential $\mu_{n,\text{sw}}$ is obtained in a corresponding manner.

\subsection{Model for Equilibrium}\label{ssc::Model-Equ}

A pure liquid (i.e.\ one-component system with no nanoparticles) forms a spherically shaped droplet that evaporates due to the mechanical imbalance caused by the Laplace pressure of the droplet with radius $R$. Even if we are at bulk liquid-vapor coexistence, i.e.\ where mechanical and chemical equilibrium are in principle fulfilled, for a convex body such as a droplet, mechanical equilibrium does not hold. Rather, we have
\begin{equation}\label{eq::laplace-pressure}
	P_\text{in}-P_\text{out}=\frac{2\gamma}{R},
\end{equation} 
where $\gamma$ is the surface tension, $P_\text{in}$ and $P_\text{out}$ are the pressure inside and outside the droplet, respectively. Thus, mechanical equilibrium is violated; only for very large droplets are the pressures approximately equal. Hence, the droplet will go through an evaporation process that accelerates with decreasing radius $R$, since the Laplace pressure increases.

In contrast, nanoparticle-laden droplets can exhibit a mechanical equilibrium by compensating the effect of the Laplace pressure. The attractive forces between the nanoparticles and the liquid can lower the pressure inside the droplet so that mechanical equilibrium is reached.

\begin{figure}
\includegraphics[scale=0.3]{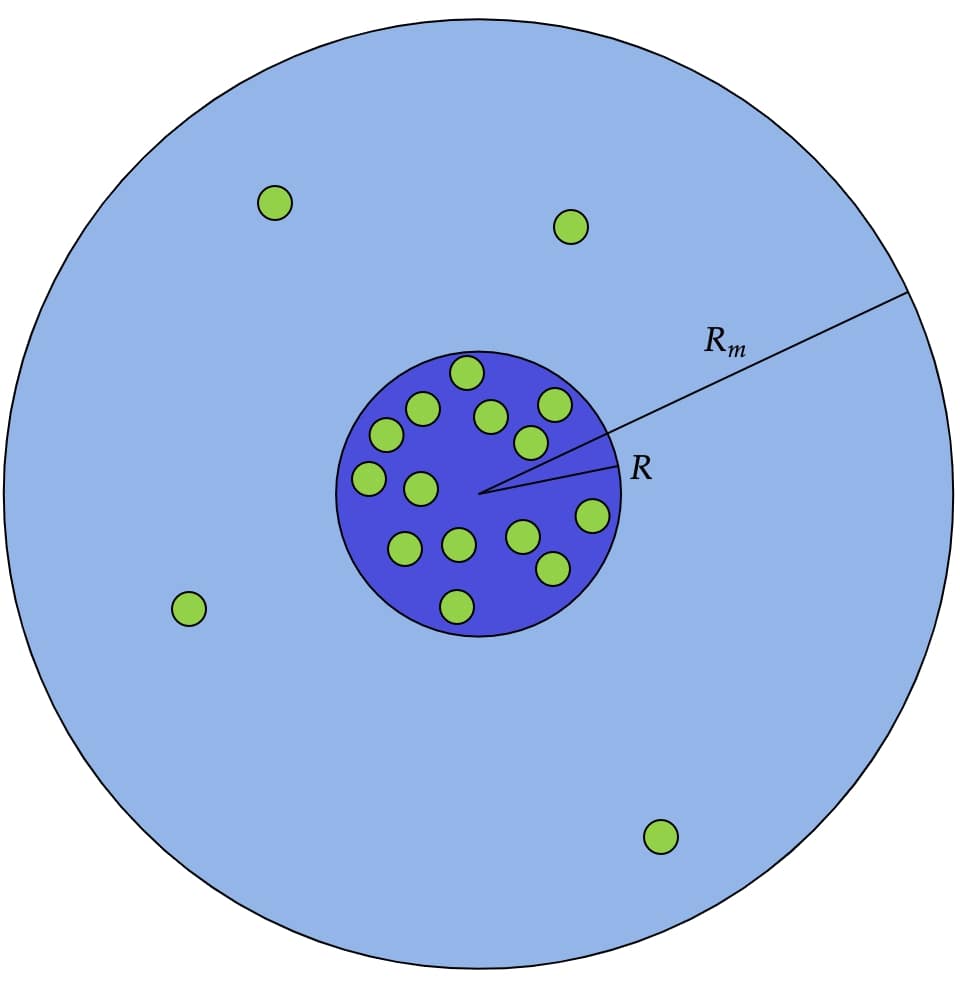}
\caption{Schematic picture of a nanoparticle-laden droplet in a spherical system of radius $R_m$ containing a droplet (dark blue) of radius $R$ in equilibrium surrounded by vapor (light blue). A fraction $\xi$ of the nanoparticles (green) are inside the droplet.\label{fig::model_droplet}}
\end{figure}

Let us consider a spherical system of radius $R_m$ containing a droplet of size $R<R_m$ with a liquid packing fraction $\eta_l$ surrounded by the vapor with packing fraction $\eta_v$, as illustrated in Fig.~\ref{fig::model_droplet}. We furthermore assume a configuration of nanoparticles of number $N_n$ in which a fraction $\xi$ of these particles are evenly distributed inside the droplet and therefore $1-\xi$ outside. Then, we can easily find the (uniform) nanoparticle packing fractions inside and outside of the droplet
\begin{equation}\label{eq::packing-fractions-nano}
	\eta^{(\text{in})}_n=\frac{\xi N_n}{8\tilde{R}^3},\quad \eta^{\text{(out)}}_n=\frac{(1-\xi)N_n}{8q^3(\tilde{R}_m^3-\tilde{R}^3)},
\end{equation}
where we use the dimensionless variables $\tilde{R}=R/\sigma_l$, $\tilde{R}_m=R_m/\sigma_l$ and size ratio $q=\sigma_l/\sigma_n$.

Mechanical equilibrium then occurs if we have
\begin{equation}\label{eq::mech-equ-nano}
	\tilde{P}(\eta^{(\text{in})}_l,\eta^{\text{(in)}}_n) = \tilde{P}(\eta^{(\text{out})}_l,\eta^{\text{(out)}}_n)+\frac{\pi\tilde{\gamma}}{3\tilde{R}},
\end{equation}
where $\tilde{\gamma}$ is defined as in Sec.~\ref{ssc::sw-fluid} and we have introduced the reduced pressures $\tilde{P}=\beta\frac{\pi}{6}\sigma_l^3P$ of the binary mixture given in Eq.~\eqref{eq::total-pressure}. Of course, the presence of nanoparticles will affect the surface tension \cite{Melih2025effective}, therefore leading to a different value than found from the pure liquid system. For big droplets ($R\gg\sigma_l$), however, the contribution stemming from the surface in Eq.~\eqref{eq::mech-equ-nano} will be negligible. Only for small-scale droplets, 10 to 20 times the size of liquid particles, does the influence associated to the surface contribution become important.
Besides the mechanical equilibrium, we also have to take into account the chemical equilibria of the liquid and the nanoparticles. For the former, we impose a chemical potential $\mu_0$ which is less than the chemical potential at liquid-vapor coexistence, $\mu_c$, thus a relative humidity $H_r$ less than 100\%. Hence, we have
\begin{align}\label{eq::chem-equilibria}
	\mu_0 &= \mu_l(\eta^{(\text{in})}_l, \eta^{(\text{in})}_n)\\\nonumber
	\mu_n(\eta^{(\text{in})}_l,\eta^{(\text{in})}_n) &= \mu_n(\eta^{(\text{out})}_l,\eta^{(\text{out})}_n).
\end{align}
With the first line in Eq.~\eqref{eq::chem-equilibria} we impose the same chemical potential inside the droplet. 
It is crucial to mention that the equilibrium laden droplet size will depend on the size $R_m$ of the system as long as $\xi<1$. This can be pictured in the following way: If we have found an equilibrium droplet with a given size where a fraction $\xi<1$ is inside the droplet, then due to chemical equilibrium there must be a certain amount ($1-\xi$) of nanoparticles outside the droplet in the gaseous phase. Now, if we expand the system size, keeping the number of nanoparticles $N_n$ constant, there is additional volume outside the droplet that has to be filled by the nanoparticles in order to maintain chemical equilibrium. Therefore, the droplet itself has less nanoparticles available for stabilization and consequently has to shrink. The only possibility of a system-independent configuration is the case where $\xi\approx 1$, i.e. (almost) all nanoparticles are inside the droplet.

We want to emphasize that the model outlined above is equivalent to the capillarity model discussed in \cite{Archer2023}. Assuming a system of volume $V$ consisting of pure vapor of pressure $P_v$, the change in grand potential $\Delta\Omega$ due to the insertion of a droplet of radius $R$ and pressure $P_d$ is given by
\begin{equation}\label{eq::change-grand-pot}
    \Delta\Omega(\rho_l,R) = -\frac{4}{3}\pi R^3\left(P_d(\rho_l, R)-P_v\right)+4\pi R^2\gamma
\end{equation}
from which the equilibrium droplet radius arises through the condition
\begin{equation}\label{eq::cond-drop-radius}
    \frac{\partial\Delta\Omega(\rho_l,R)}{\partial R}=0.
\end{equation}
Furthermore, imposing a chemical potential $\mu$ close to the coexistence of the solvent, the equilibrium radius $R$ and the liquid density of the droplet $\rho_l$ can be calculated. At this point, it is important to note that Eq.~\eqref{eq::mech-equ-nano} is equivalent to Eq.~\eqref{eq::cond-drop-radius} of the capillarity model.

Given Eqs.~\eqref{eq::mech-equ-nano} and \eqref{eq::chem-equilibria} we can predict a stable droplet. Providing a system size $R_m$, an equilibrium radius $\tilde{R}$ of the droplet and a relative humidity $H_r$ (via $\mu_0$) we solve for the amount of nanoparticles $N_n$ in the system, the liquid packing fraction $\eta_l$ and the fraction of nanoparticles, $\xi$ inside the droplet. Of course, we can switch between these variables, e.g. we can, instead of fixing the inter-component energy $\epsilon_{ln}$, fix $\xi$ and ask what $\epsilon_{ln}$ is needed for that configuration. This latter case is especially important when it comes to droplets that are system-independent.

Therefore, we utilize the calculated thermodynamic quantities in order to perform the droplet minimization for a SW binary mixture. Especially in the case of a highly asymmetric binary mixture, providing a good first guess at the initialization stage is crucial for a stable minimization. For instance, by calculating the equilibrium droplet radius $R$ for a given set of SW parameters, we initialize, as mentioned in Sec.~\ref{ssc::laden-droplet}, the density profiles $\rho_l$ and $\rho_n$ accordingly together with the right amount of nanoparticles, $N_n$, necessary for droplet stabilization. If, for example, we started at a droplet radius differing from the equilibrium one, the corresponding minimization process would first of all need more time to equilibrate and also be much more prone to become unstable. Hence, by providing suitable quantities for initialization and employing several minimization stages we render the overall minimization process feasible.

\section{Results and Discussion}\label{sc::results}

\begin{figure}
\includegraphics[scale=0.05]{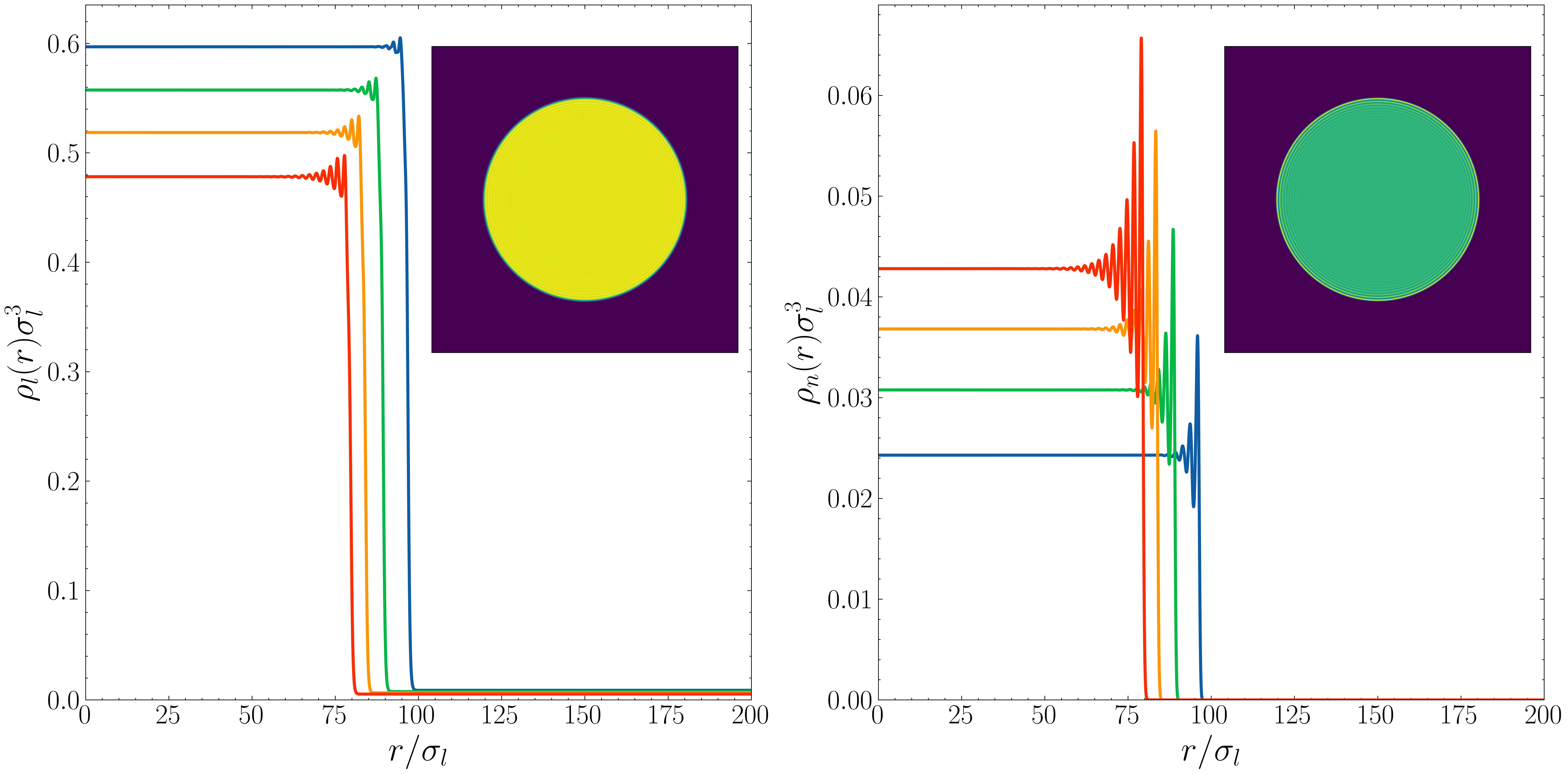}
\caption{Density profiles $\rho_l(r)$ (left) and $\rho_n(r)$ (right) of a 2:1 droplet with $\beta\epsilon_{nn}=0,\,\beta\epsilon_{ln}=3.78,\,\lambda_{nn}=1.01$ and $N_n=92281$. For $H_r=50\%$ (red) the droplet has a radius $\tilde{R}=80$ which increases as the humidity is raised to $H_r=60\%$ (green), $H_r=70\%$ (orange) and $H_r=80\%$ (blue). The insets show a heatmap plot of the density in a 2:1 droplet for $H_r=50\%$. \label{fig::2_to_1}}
\end{figure}

Having outlined the model for nanoparticle-laden stable droplets in equilibrium, see Sec.~\ref{ssc::Model-Equ}, we can apply the numerical scheme presented in Sec.~\ref{ssc::laden-droplet} in order to obtain radially symmetric density profiles $\rho_l(r)$ and $\rho_n(r)$ of the liquid and nanoparticles. For the liquid, we fix the associated SW parameters as $\beta\epsilon_{ll}=1.2$ and $\lambda_{ll}=1.5$, leading to coexistence densities or packing fractions, given in Eq.~\eqref{eq::coex-etas}. Adding nanoparticles into the system can stabilize a droplet of certain radius in an environment of the prescribed humidity. Contrary to a more realistic model for viruses in water, where a nanoparticle is about thousand times larger than a liquid particle, we only consider the case of nanoparticles being ten times the size of the liquid particles. One main reason for this is the numerical challenge which comes along when minimizing in DFT for a highly asymmetric binary system. These droplets can be generated in different environments, i.e.\ for different values of the humidity $H_r$. Formally, the latter is defined by $H_r=100\times P(\mu_0)/P(\mu_c)$, where $\mu_0\leq\mu_c$ is the chemical potential that we choose for our calculations. 
For instance, $H_r=50\%$ corresponds to a rather comfortable moisture level given in a typical room of temperature around 20 $^{\circ}\text{C}$.

Then, by further specifying the interaction parameters $\epsilon_{nn},\,\epsilon_{ln},\,\lambda_{nn}$ and $\lambda_{ln}$ together with a proportion of nanoparticles staying inside the droplet, $\xi\approx 1$, we can predict the equilibrium radius $\tilde{R}=R/\sigma_l$ of the droplet and the amount $N_n$ of nanoparticles in the system necessary to stabilize the droplet. The size ratio $q=\sigma_l/\sigma_n$ takes into account the possibility of having an asymmetric binary mixture. If for instance $q=1/2$, then we refer to the corresponding droplet as a 2:1 droplet.

\begin{figure}
\includegraphics[scale=0.05]{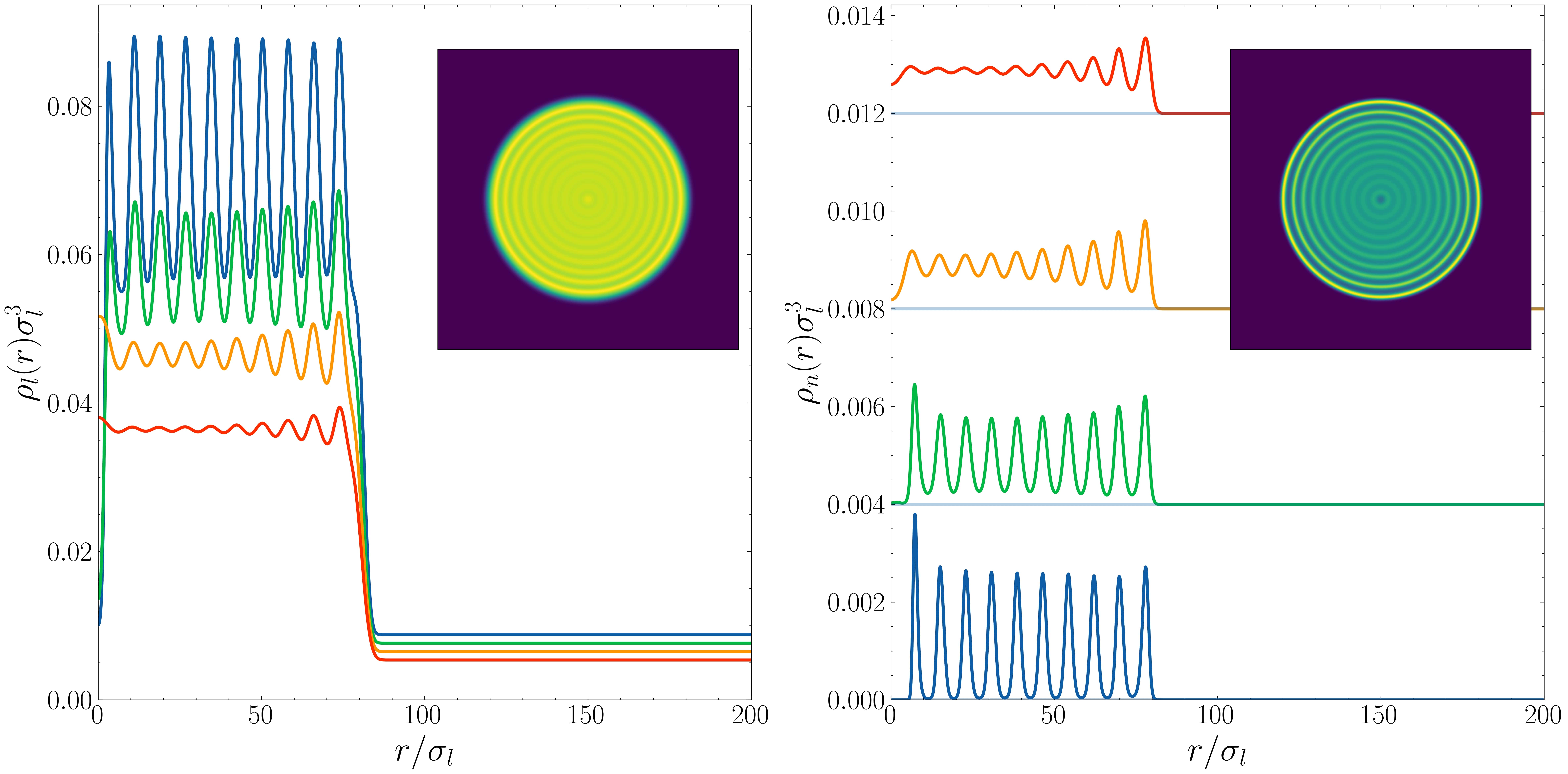}
\caption{Density profiles $\rho_l(r)$ (left) and $\rho_n(r)$ (right) of a 10:1 droplet with $\beta\epsilon_{nn}=0,\,\beta\epsilon_{ln}=3.69,\,\lambda_{nn}=1.01$ and $N_n=1983$ with an offset represented by blue lines. For $H_r=50\%$ (red) the droplet has a radius $\tilde{R}=80$, which increases as the humidity is raised to $H_r=60\%$ (green), $H_r=70\%$ (orange) and $H_r=80\%$ (blue). The insets show a heatmap plot of the densities of the 10:1 droplet for $H_r=50\%$. \label{fig::10_to_1}}
\end{figure}

We present scenarios where we set $\epsilon_{nn}=0$ while $\epsilon_{ln}<0$ and also scenarios where the mixing rule of the energy, i.e. $\epsilon_{ln}=\sqrt{\epsilon_{ll}\epsilon_{nn}}$ is satisfied. The corresponding mixing rule for the interaction range, i.e. $\lambda_{ln}=(\lambda_{ll}\sigma_l+\lambda_{nn}\sigma_n)/(\sigma_l+\sigma_n)$ is obeyed throughout the discussion.
We expect the interaction parameter $\epsilon_{ln}$ to have the most impact on droplet stabilization as it describes the solubility of nanoparticles inside the liquid while the interaction energy between the nanoparticles $\epsilon_{nn}$ becomes important only for sufficiently high nanoparticle densities, when nanoparticles start to interact with each other. \textcolor{black}{The specific values we use for $\epsilon_{ln}$ or $\epsilon_{nn}$ are bounded by two constraints: On the one hand, we want to prevent the nanoparticles from building clumps within the droplet, i.e.\ the interaction strengths $\epsilon_{ln}$ and $\epsilon_{nn}$ must not be too large. In addition, the solubility of the nanoparticles inside the droplet is only controlled by the inter-component interaction strength $\epsilon_{ln}$, which must be large enough to let the nanoparticles prefer to be inside the droplet. The values for $\epsilon_{ln}$ and $\epsilon_{nn}$ we employ here respect these constraints.}

Before presenting the case of 10:1 size ratio, it is also interesting to first regard the case of a less asymmetric case of 2:1. In Fig.~\ref{fig::2_to_1} we show density profiles of a droplet having a radius $\tilde{R}=80$ at a humidity $H_r=50\%$ with the corresponding SW interaction parameters given in the figure caption. We observe that inside the droplet the density profiles of liquid and nanoparticles are constant, exhibiting oscillations only close to the surface of the droplet. \textcolor{black}{These oscillations become more pronounced as we lower humidity. The typical appearance of the oscillatory behavior at liquid-vapor interfaces can be explained through the Fisher-Widom line separating pure exponential from exponentially damped decay of density distributions of SW fluids \cite{evans1993asymptotic, brader2001entropic}}. Further we see that the liquid density $\rho_l(r)$ increases as the the humidity is raised up while the nanoparticle density $\rho_n(r)$ diminishes. This is, of course, due to the fact that as the droplet grows the nanoparticles have more space available.

\begin{figure}
\includegraphics[scale=0.17]{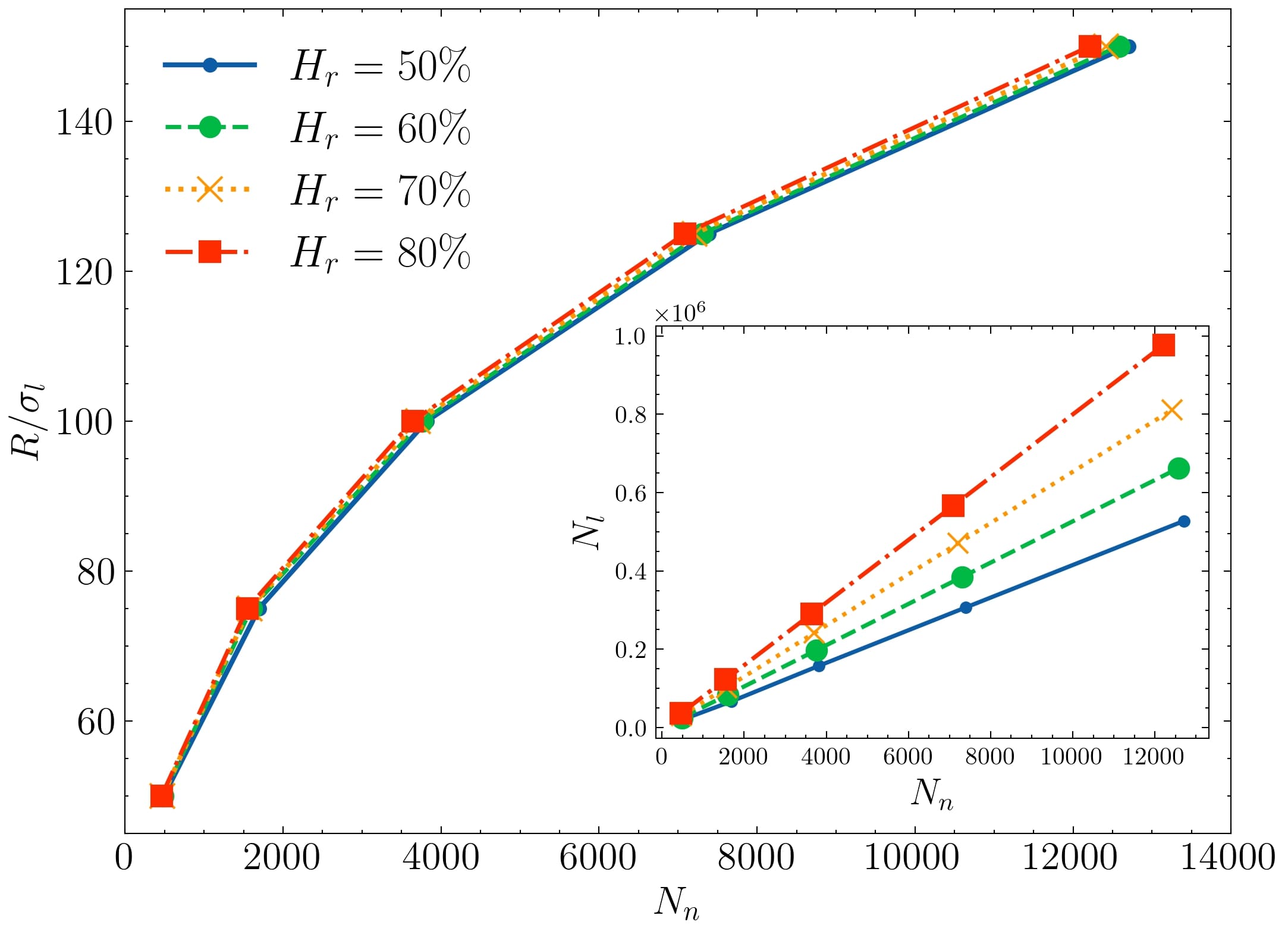}
\caption{The droplet radius $R$ is plotted as a function of nanoparticle number $N_n$ with a fixed interaction strength $\beta\epsilon_{ln}=3.69$ of Fig.~\ref{fig::10_to_1} for several humidity values considered here. The inset displays the dependence of the number of liquid particles $N_l$ on the nanoparticle number $N_n$.\label{fig::R_Nn}}
\end{figure}

Figure~\ref{fig::10_to_1} shows a 10:1 droplet with the same radius as the 2:1 droplet in Fig.~\ref{fig::2_to_1}. As the size ratio now is increased, the amount of nanoparticles needed for stabilization is much less. Furthermore, we observe much stronger oscillations, not only close to the surface but also inside the droplet, especially for high humidity.  What we can also observe is that the droplet radius is almost unchanged, in contrast to the 2:1 droplet demonstrated in Fig.~\ref{fig::2_to_1}. The oscillations, clearly visible in the insets of Fig.~\ref{fig::10_to_1}, have a pattern of concentric circles for both the liquid and nanoparticles in such a way that a maximum of the former meets a minimum of the latter. Hence, the particles have a tendency to align themselves in shells which becomes stronger as the humidity is increased, i.e.\ there are more liquid particles in the droplet thus enforcing the nanoparticles to congregate in shells. In Fig.~\ref{fig::R_Nn} we present the model predictions on the droplet radius $R$ and number of liquid particles inside the droplet $N_l$ as a function of the number of nanoparticles $N_n$. For that, we take into account several values of humidity $H_r$. By adding nanoparticles into the droplet, the corresponding droplet radius increases according to the expected law $R\propto N_n^{1/3}$. 
%\bg{Should we put this on a log plot so we can see the power law scaling?}.
% Andy: No. Melih is right to not do log-plot, since the range of N_n values is only one order of magnitude.
Matching to the observation made on Fig.~\ref{fig::10_to_1}, the droplet radius barely varies with changing humidity. However, as can be inferred from the inset of Fig.~\ref{fig::R_Nn}, the amount of liquid particles $N_l$ inside the droplet is susceptible to humidity, changing in a linear fashion with respect to $N_n$. Note that a similar observation was also made in \cite{Archer2023}, where the slope is increased for high humidity values. Discrepancies between the lattice-DFT based calculations and the present model output are believed to be caused by the use of a value for the interfacial surface tension $\gamma$ that is for the pure liquid system.

\begin{figure}
\includegraphics[scale=0.05]{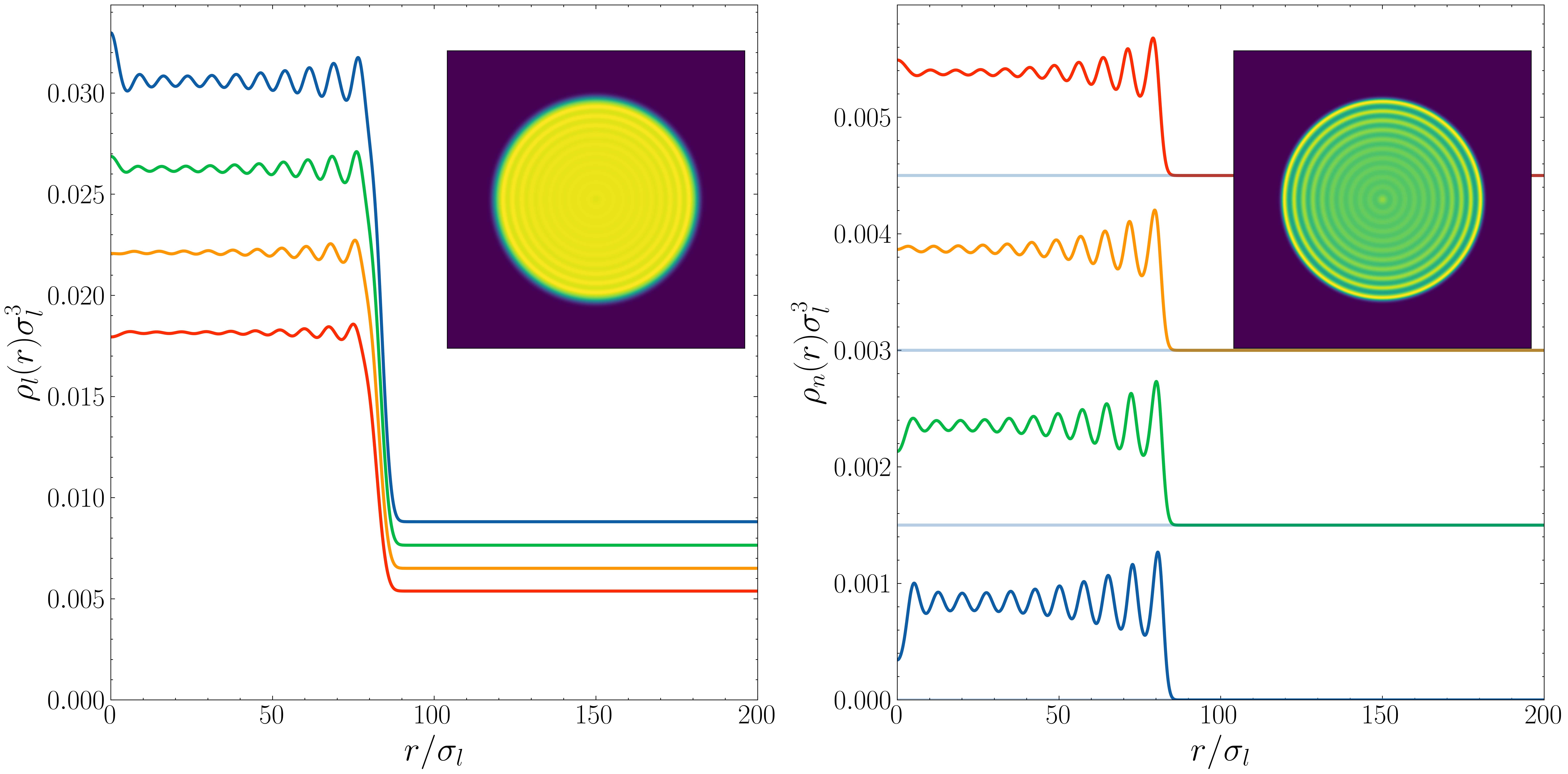}
\caption{Density profiles $\rho_l(r)$ (left) and $\rho_n(r)$ (right) of a 10:1 droplet with $\beta\epsilon_{nn}=6.96,\,\epsilon_{ln}=\sqrt{\epsilon_{ll}\epsilon_{nn}},\,\lambda_{nn}=1.01$ and $N_n=2063$ with an offset represented by blue lines. For $H_r=50\%$ (red) the droplet has a radius $\tilde{R}=80$ which increases as the humidity is raised to $H_r=60\%$ (green), $H_r=70\%$ (orange) and $H_r=80\%$ (blue). The insets show a heatmap of the 10:1 droplet for $H_r=50\%$. \label{fig::10_to_1_mixing_rule}}
\end{figure}

We can also realize a numerically challenging droplet with 10:1 size ratio that respects the mixing rule of energy, i.e. $\epsilon_{ln}=\sqrt{\epsilon_{ll}\epsilon_{nn}}$, as is shown in Fig.~\ref{fig::10_to_1_mixing_rule}. Thus, we need slightly more nanoparticles in order to keep the droplet of the same size stable compared to the case of Fig.~\ref{fig::10_to_1}, where we set $\epsilon_{nn}=0$. We also notice that the density oscillations are less pronounced throughout the droplet, both for the liquid and nanoparticles. In addition, the density profiles of the nanoparticles have almost the same magnitude and shape between each other, only showing differences in the center of the droplet. Comparing the overall magnitudes of the liquid densities inside the droplet of Fig.~\ref{fig::10_to_1} and Fig.~\ref{fig::10_to_1_mixing_rule} we see that with the mixing rule for the energy (i.e.\ with attractive interactions between the nanoparticles), the liquid densities inside the droplet are much lower. This is because with the mixing rule we obtain $\beta\epsilon_{ln}\approx 2.89$ which is lower than the value used in Fig.~\ref{fig::10_to_1} and therefore is a weaker attraction between liquid and nanoparticles. The latter simply does not need as many liquid particles as in Fig.~\ref{fig::10_to_1} due to the additional attraction between the nanoparticles necessary to keep the droplet stable.
Furthermore, as is observed for the droplet of Fig.~\ref{fig::10_to_1}, the radius of the droplet of Fig.~\ref{fig::10_to_1_mixing_rule} varies only marginally.

We conclude that it is possible to minimize a SW binary mixture within the framework of DFT in order to obtain structural information. The main feature of these density profiles is the appearance of oscillations close to the surface of the droplet. These become more pronounced as we increase the size ratio, $q$, of the SW binary mixture, see Fig.~\ref{fig::2_to_1} and Fig.~\ref{fig::10_to_1}. For the 2:1 droplet, these oscillations solely appear close to the surface, becoming stronger as the humidity is decreased. Here, it is important to note that the liquid density inside the droplet is close to the density at coexistence of the pure liquid. However, considering the 10:1 droplet, the corresponding liquid density inside the droplet is lowered, especially in the case of respecting the energy mixing rule, $\epsilon_{ln}=\sqrt{\epsilon_{ll}\epsilon_{nn}}$. Furthermore, we observe oscillations throughout the droplet, which become stronger in a humid environment where the droplet radius remains constant.
\textcolor{black}{The work of \cite{jin2023morphology} using MD techniques  studies the diameter and morphological properties of droplets that, when initially containing nanoparticles, undergo an evaporation process with several outcomes prescribed by the Peclet-number. In particular, crust formation is observed for a given range of the Peclet-number, which resembles our findings in Fig.~\ref{fig::10_to_1} and Fig.~\ref{fig::10_to_1_mixing_rule}. Furthermore, we find in \cite{jin2023morphology} snapshots of density profiles of solvent and solute particles that exhibit oscillatory behavior near the surface of the droplet. Similar observations can also be made in \cite{chen2013molecular}. Note that we only present profiles at equilibrium whereas the results from MD simulations are fully dynamical.}

\begin{figure}
\includegraphics[scale=0.17]{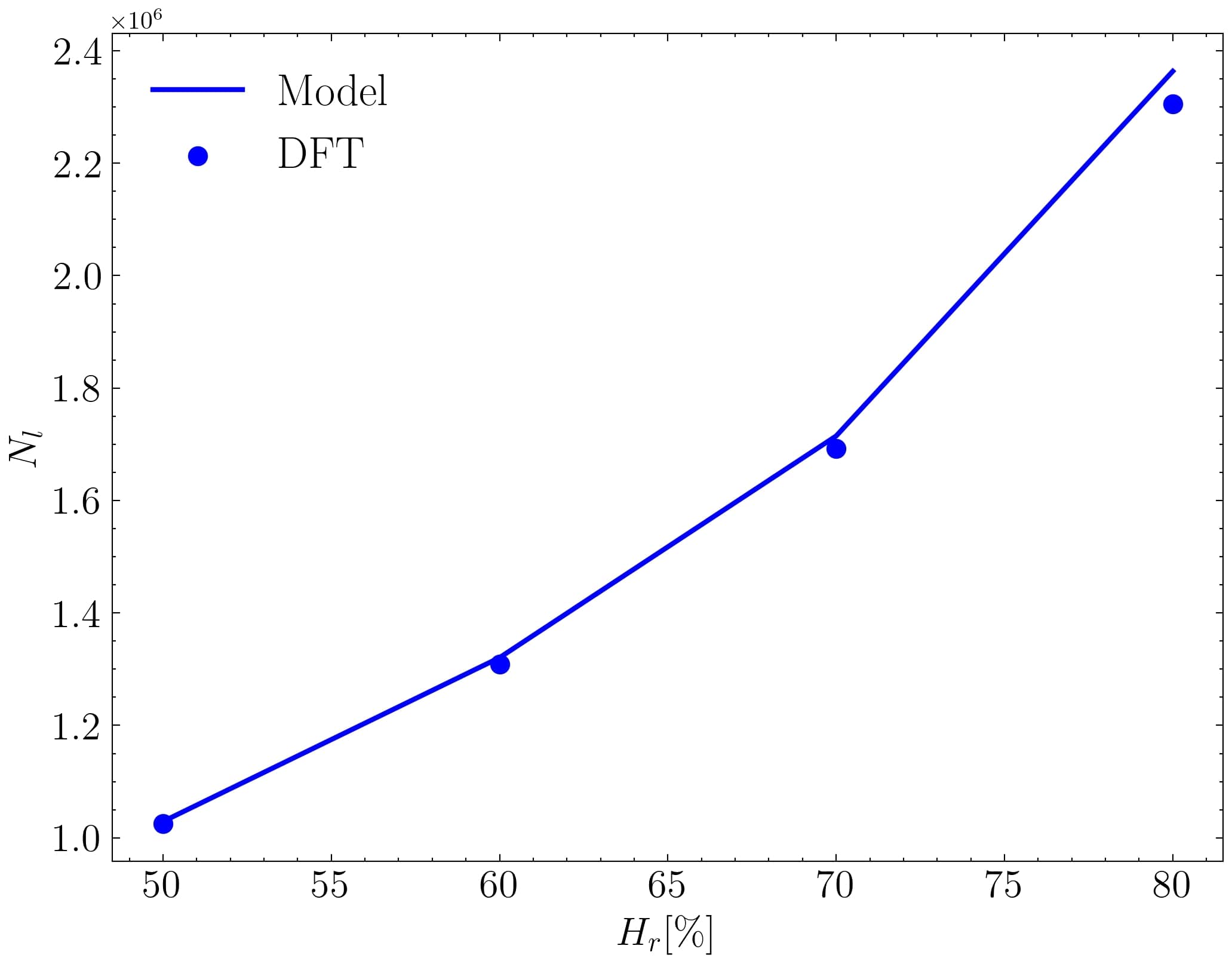}
\caption{The number of liquid particles $N_l=\frac43\pi R^3\rho_l$ inside the droplet of the 2:1 system of Fig.~\ref{fig::2_to_1} plotted against humidity $H_r$.\label{fig::nl_2_1}
}
\end{figure}

\begin{figure}
\includegraphics[scale=0.17]{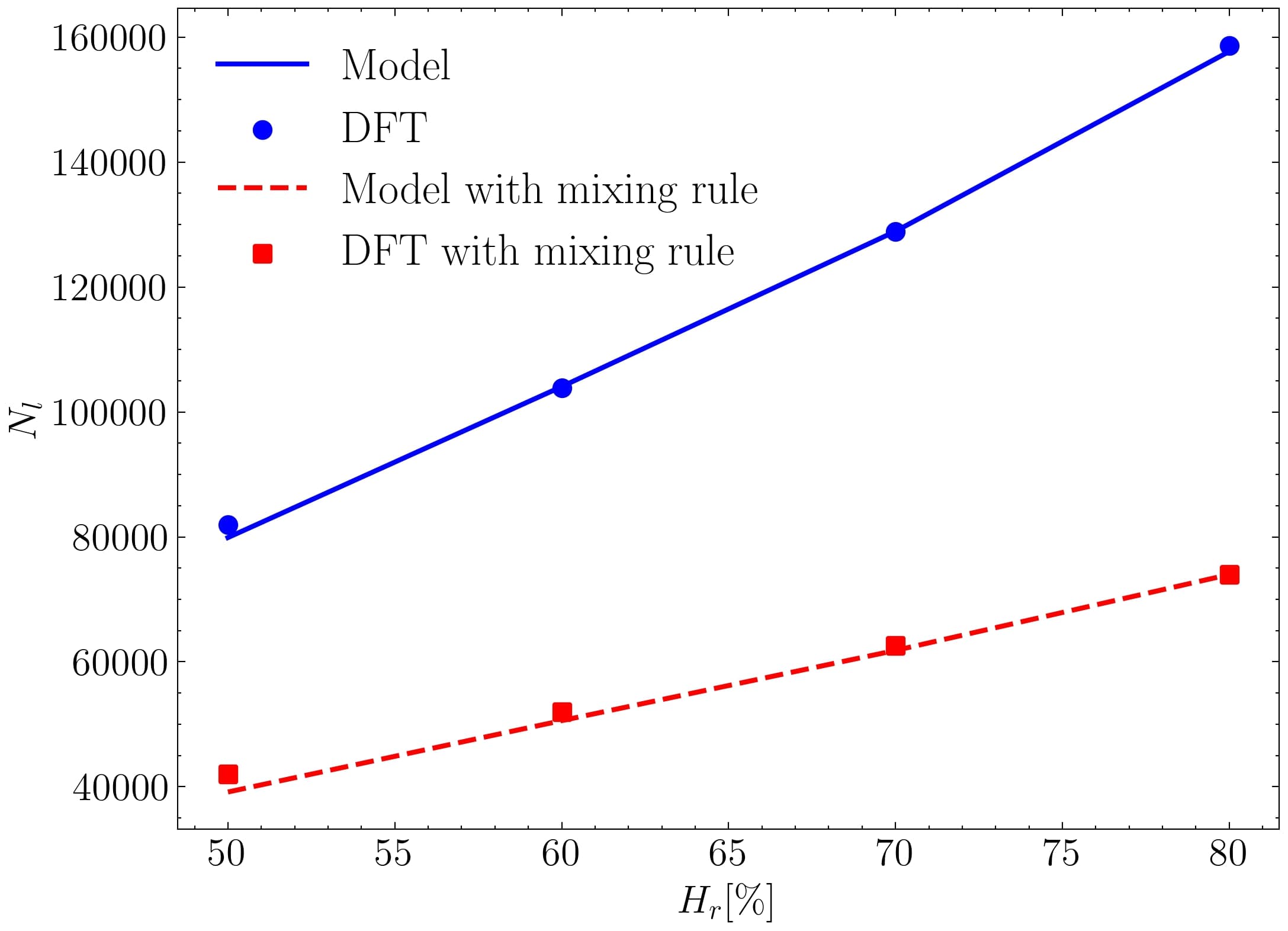}
\caption{The number of liquid particles $N_l=\frac43\pi R^3\langle\rho_l\rangle$ inside the droplet of the 10:1 system of Fig.~\ref{fig::10_to_1} and Fig.~\ref{fig::10_to_1_mixing_rule} plotted against humidity $H_r$.\label{fig::nl_10_1}}
\end{figure}

Similarly to the discussion in \cite{Archer2023}, where the amount of liquid particles inside the corresponding droplet was compared to the prediction of the capillarity model, we want to consider the change of liquid particles by altering humidity. Here, our model of Sec.~\ref{ssc::Model-Equ} predicts the uniform liquid packing fraction $\eta^{(\text{in})}_l$ inside the droplet of radius $R$. Therefore, the amount of liquid particles $N_l$ is given by
\begin{equation}\label{eq::nl-model}
    N_l = 8\eta^{(\text{in})}_l \tilde{R}^3
\end{equation}
where again $\tilde{R}$ is the droplet radius in units of liquid diameter $\sigma_l$.
On the other hand, the corresponding DFT calculations provide spherically symmetric equilibrium density profiles $\rho_l(r)$ from which we can obtain the number of liquid particles inside the droplet
\begin{equation}\label{eq::nl-dft}
    N_l = \frac{4}{3}\pi R^3\langle\rho_l\rangle
\end{equation}
with the spatially averaged density
\begin{equation}\label{eq::rho-averaged}
    \langle\rho_l\rangle = \frac{3}{R^3}\int_0^R\text{d}r\,r^2\rho_l(r).
\end{equation}
Thus, the model prediction Eq.~\eqref{eq::nl-model} can be compared to the DFT calculation Eq.~\eqref{eq::nl-dft} of $N_l$ for several solutions of the droplet. Fig.~\ref{fig::nl_2_1} shows the amount of liquid particles $N_l$ of the 2:1 system given in Fig.~\ref{fig::2_to_1} for several values of humidity $H_r$ that we investigate here. We can observe that the DFT predictions are slightly below the values of the model which becomes more visible towards higher humidity values. This disparity can be explained by the fact that the DFT calculation of Eq.~\eqref{eq::nl-dft} needs a specific value for the radius $R$ that must be inferred from the density profile. One possibility to define the droplet radius is to look where the transition to the vapor density within some threshold occurs. For an uncertainty $\delta r$ of the droplet radius, the corresponding relative error of the volume is of the order $3\delta r/R$. In our cases we typically have $|\delta r|\leq2\sigma_l$, hence a relative error of approximately 8\%. The largest deviation of Fig.~\ref{fig::nl_2_1} is ca. 4.4\% and is therefore within the uncertainty of the DFT calculation. Finally, we have employed the surface tension $\tilde{\gamma}=0.25$ of the pure liquid system which, due to the presence of solute particles in the system, is altered. Hence, this deviation in surface tension will induce differences in the equilibrium profiles of the liquid and nanoparticles and thus slightly modified radii.

In accord with our expectation, the number of liquid particles inside the droplet increases with ascending humidity. The magnitude of $N_l$ is around $10^6$ which is clear from the high values of the density profiles in Fig.~\ref{fig::2_to_1}. In the same way, Fig.~\ref{fig::nl_10_1} displays the number of liquid particles $N_l$ of the 10:1 system for both scenarios presented in Fig.~\ref{fig::10_to_1} and Fig.~\ref{fig::10_to_1_mixing_rule}. Firstly, we see good consistency between the model and DFT predictions, with the highest deviation of ca. 6.8\%. Furthermore, the scale of $N_l$ is of the order ten times smaller to the case of the 2:1 system. Finally, the increase of $N_l$ follows a linear shape, in contrast to Fig.~\ref{fig::nl_2_1} where the increase of $N_l$ is parabolic.

\section{Conclusion}\label{sc::conclusion}
Our work presented here shows that a simple thermodynamic model of a droplet containing solute particles differing in size can provide a theoretical framework for determining the equilibrium droplet size at given humidity; see Sec.~\ref{ssc::Model-Equ}. We furthermore employed continuum DFT calculations in Sec.~\ref{sc::results} that yield density profiles of solvent and solute particles exhibiting interesting structures inside the droplet. While for a smaller size ratio of 2:1 between solute and solvent particle the corresponding profiles only have structure close to the surface of the droplet, for a size ratio of 10:1 oscillations occur throughout the interior of the droplet, as can be seen in Fig.~\ref{fig::10_to_1} and Fig.~\ref{fig::10_to_1_mixing_rule}. Also, the liquid density needed to maintain the droplet is much smaller than in the case of the 2:1 droplet, see Fig.~\ref{fig::2_to_1}. This is due to the fact that bigger solute particles need fewer liquid particles in order to equilibrate the droplet. If there are also attractive interactions between solute particles themselves, then even fewer liquid particles are necessary, as can be inferred from Fig.~\ref{fig::10_to_1_mixing_rule}. The latter is at the same time a scenario where the Lorentz-Berthelot mixing rules are satisfied.

Given the density profiles of a DFT calculation, we can make a comparison between the model output on the predicted liquid density inside the droplet, or equivalently, the number of liquid particles contained in the droplet and the result obtained from the minimized density profile. Fig.~\ref{fig::nl_2_1} and Fig.~\ref{fig::nl_10_1} display the amount of liquid particles as a function of humidity. The agreement between the thermodynamical model outlined in Sec.~\ref{ssc::Model-Equ} and the DFT calculations is convincing, particularly given the fact that DFT calculations are crucial to understand the structures of the densities.

In summary, we can state that a stable laden droplet can be realized within the framework of classical DFT by using several (more sophisticated) minimization procedures than are usually performed. Due to the limits of FMT to account for highly asymmetric mixtures of hard-spheres with possible attractive forces, we restricted our analysis to the case of 10:1 size ratio between solutes and solvents. Although the DFT density profiles exhibit a lot of structure in the interior of the droplet, i.e.\ deviating clearly from the uniform distribution of the model, we nonetheless could find good agreement between the thermodynamic model and DFT.

We have demonstrated that the presence of nanoparticles in the droplet stabilises them against complete evaporation. This fact can explain the long lifetime of virus laden aerosols.


%\bibliography{laden_droplet.bib}

\end{document}